\documentclass[%
 reprint,
 amsmath,amssymb,
prb,
]{revtex4-2}

\usepackage{graphicx}
\usepackage{dcolumn}
\usepackage{bm}
\usepackage{color,soul}
\usepackage{comment}

\begin{document}

\preprint{APS/123-QED}

\title{Emergent Classical Spin Liquid Phases in an Ising Lattice via Size Effects}

\author{Pratik Brahma$^{1}$, Sayeef Salahuddin$^{1,2}$}
\affiliation{
$^{1}$ Department of Electrical Engineering and Computer Sciences, University of California, Berkeley\\
$^{2}$ Materials Science Division, Lawrence Berkeley National Laboratory
}
\email{pratik\_brahma@berkeley.edu; sayeef@berkeley.edu}

\begin{abstract}

We show that a classical spin liquid phase can emerge from an ordered magnetic state in the two-dimensional frustrated Shastry-Sutherland Ising lattice due to lateral confinement. Two distinct classical spin liquid states are stabilized (i) long-range spin-correlated dimers, and (ii) exponentially decaying spin-correlated disordered states, depending on widths of W=3n, 3n+1 or W=3n+2, n being a positive integer. Stabilization of spin liquids in a square-triangular lattice moves beyond the conventional geometric paradigm of kagome, triangular or tetrahedral arrangements of antiferromagnetic ions, where spin liquids have been discussed conventionally. 

\end{abstract}

\maketitle

Geometric frustration is an essential tool for stabilizing emergent quantum phenomena in magnetic materials due to the absence of a unique state that simultaneously satisfies all interactions among the magnetic ions. 
The quantum spin liquid state is an example of a frustration-driven emergent phase comprising highly correlated spin fluctuations down to zero temperature \cite{Broholm2020, Zhou2017, Savary2017, Balents2010}. 
This disordered state harbors many exotic quantum properties like fractional charges, and long-range entanglement, which is relevant for high-temperature superconductivity \cite{Anderson1987} and topological quantum computing \cite{Ioffe2002}.
Similarly, the classical counterpart defined by classical magnetic moments on a geometrically frustrated lattice demonstrates finite thermodynamic entropy per spin down to zero temperature without any quantum entanglement \cite{Myers2000, Gingras2011, Ramirez1999}.
This is due to the presence of macroscopic ground state degeneracy, which is considered a possible route to quantum spin liquids \cite{Niggemann2020}.

As yet, these exotic phases of matter have been stabilized only in a few select geometric arrangements: (i) two-dimensional Ising triangular lattice \cite{Arh2022, Li2015}, (ii) two-dimensional kagome lattice \cite{Kermarrec2021}, (iii) three-dimensional hyper-hyper kagome lattice \cite{Chillal2020}, and (iv) three-dimensional pyrochlore spin ice \cite{Carrasquilla2015, Ramirez1999}. In this context, geometric confinement may provide a new and powerful way to induce novel phases. Indeed, in the recent years, confinement has been utilized for enhanced ferroelectricity in unit cell thick HfO$_2$ and ZrO$_2$\cite{Cheema2020, Cheema2022}, ferromagnetic order in bilayer van der Waals crystal Cr$_2$Ge$_2$Te$_6$ \cite{Gong2017},high-temperature superconductivity in monolayer Bi$_2$Sr$_2$CaCu$_2$O$_{8 + \delta}$ \cite{Yu2019}. Here, we show that geometric confinement in a Sastry-Sutherland Ising lattice can lead to two distinct Spin Liquid phases, that has not been previously reported for a square-triangular lattice.

\begin{figure}
\includegraphics{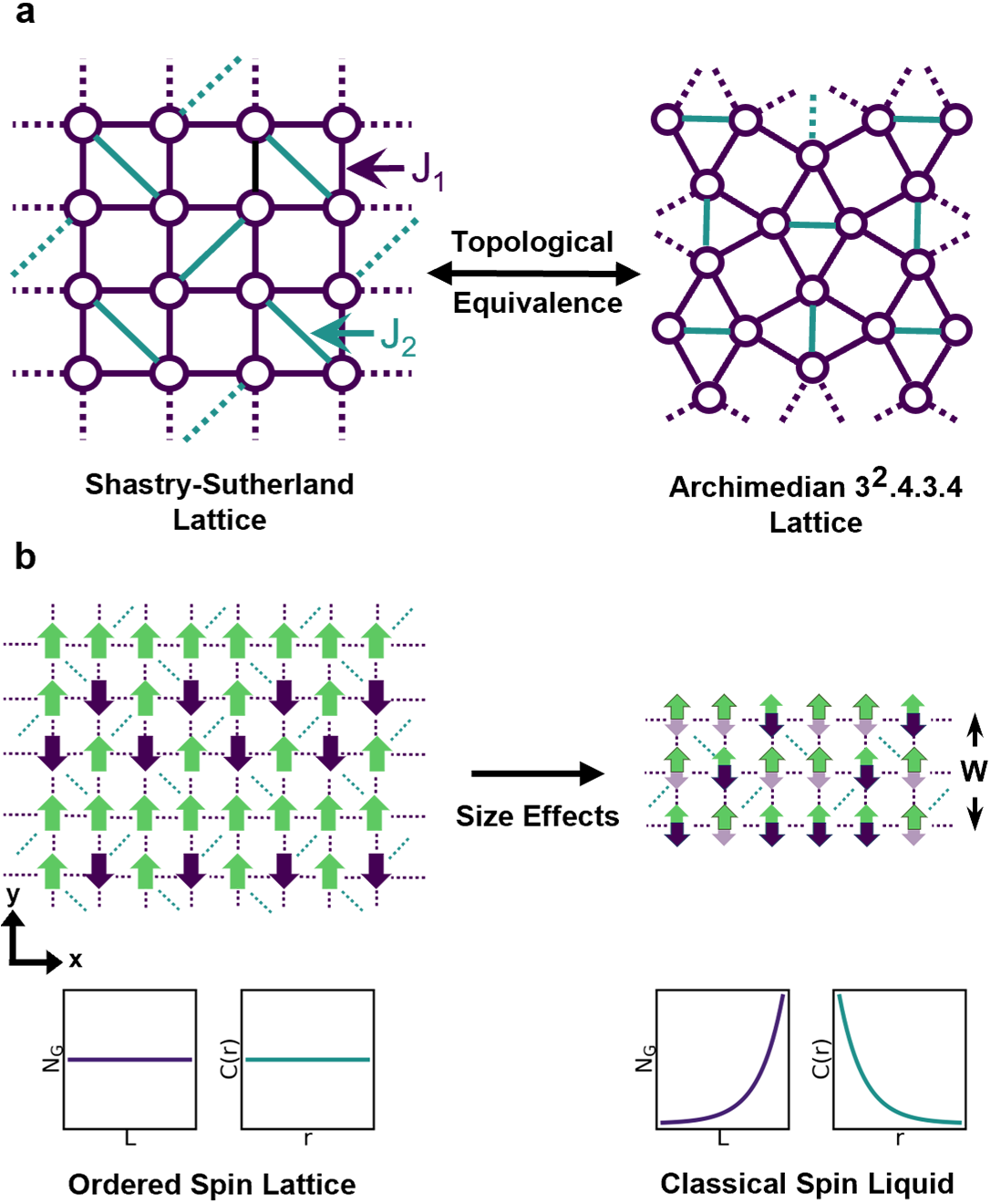}
\caption{\textbf{Emergence of classical spin liquid phase due to size effects:} (\textbf{a}) All neighboring interactions in the Shastry-Sutherland (SS) Ising lattice which is topologically equivalent to the square-triangular lattice (Archimedean $3^2.4.3.4$ lattice) \cite{Dublenych2012, Their2017}. (\textbf{b}) Introducing size effects along the lateral direction in the SS lattice via open boundary conditions. The ordered UUD spin configuration converts to a disordered classical spin liquid characterized by the presence of macroscopically degenerate ground states. The bottom graphs show the general behavior of $N_G$ (number of degenerate ground states), and $C(r)$ (spin-spin correlation) for the ordered spin state and the disordered spin liquid.  
}
\label{Fig1}
\end{figure}

We study the classical 2D frustrated Shastry-Sutherland (SS) Ising lattice (Fig.\ref{Fig1}a), which is topologically equivalent to the square-triangular Archimedean lattice \cite{Dublenych2012, Their2017}.
This lattice model and its variations may explain the presence of several fractional magnetization plateaus in quasi-two-dimensional magnetic materials, such as the rare-earth tetraborides ($\text{RB}_4$, $\text{R} = {\text{Tm}, \text{Er}, \text{Ho}}$ \cite{Yoshii2008, Kim2009, Trinh2018, Siemensmeyer2008}), where each magnetic atom in a layer is located on a 2D lattice that is topologically equivalent to the SS lattice.
Due to the large magnetic moments of the ions and crystal field effects, these compounds are well described by the classical Ising version of the SS Hamiltonian given by:
\begin{eqnarray}
\mathcal{H} \left( \mathbf{\sigma}\right) = J_1\sum_{\langle i,j \rangle} \sigma_i\sigma_j + J_2\sum_{\langle \langle i,j \rangle \rangle}\sigma_i\sigma_j + h\sum_i\sigma_i
\label{SS eqn}
\end{eqnarray}
$\sigma_i \in \{+1, -1\}$, $\langle i, j \rangle$ are the nearest neighbor spin interactions and $\langle \langle i, j \rangle \rangle$ are the diagonal and off-diagonal row-wise alternating second nearest neighbor spin interactions shown in Fig.\ref{Fig1}a. 
Frustration in the lattice arises from the competition between the $J_1$ and $J_2$ antiferromagnetic interacting bonds giving rise to the 1/3 UUD (Up-Up-Down) fractional magnetization state inside a suitable range of longitudinal magnetic field ($h_z$), where each triangular motif of the lattice contains two up spins and one down spin \cite{Dublenych2012}. 
Recent studies show the presence of a narrow spin liquid phase in the bulk quantum SS model \cite{Yang2022, Kele2022} and new quantum phases in the SS magnet SrCu$_2$(BO$_3$)$_2$ under high field and pressure \cite{Shi2022} however, the role of size effects in stabilizing spin liquids in the SS Ising lattice has yet to be explored.


\begin{figure*}
\includegraphics[width=0.76\textwidth]{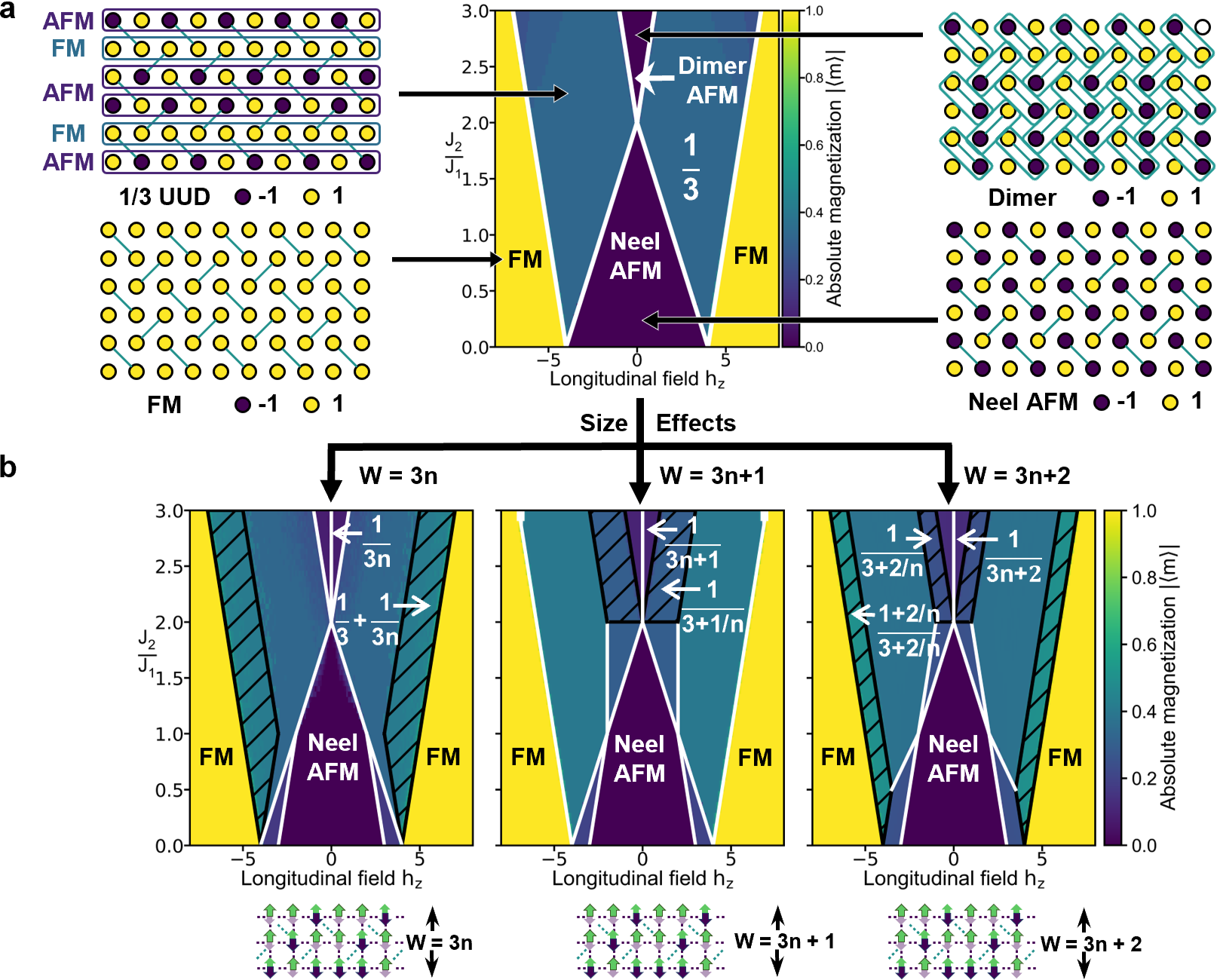}
\caption{\textbf{Transformation of magnetization phase diagram due to size effects along lateral direction:} (\textbf{a}) Magnetization phase diagram in the thermodynamic limit of the Shastry-Sutherland (SS) Ising lattice. The ground state spin configurations for all the magnetization phases are shown beside the phase diagram. Only the Dimer phase is macroscopically degenerate with zero spin correlation among the dimers (shown in green boxes), while the other spin ground states are finitely degenerate. (\textbf{b}) The phase diagrams of the laterally confined SS lattice via open boundary conditions. Depending on the width of the lattice (grouped by the modulus of 3), different regions of the 1/3 magnetization phase convert to a classical spin liquid phase (shown as black hashed regions), identified by its macroscopic degeneracy.}
\label{Fig2}
\end{figure*}

We calculate the ground state energy, ground state spin configuration, spin-spin correlation, and magnetization phase diagram of the SS Ising lattice using the total partition function \cite{Schmitz2017}. 
Since brute force calculation of the partition function is time-consuming and exponential in the lattice size, we use the transfer matrix method, which has been used for various spin lattice models \cite{Kramers1941, Onsanger1944}.
We assume the size of the lattice to be $W \times L$ and define variables and calculate the Transfer Matrix from:


\begin{eqnarray}
 T^k_{\left[\sigma_{i}\right], \left[\sigma_{i+1}\right]} &=& \exp \left(- \beta E^k_{\left[\sigma_{i:i+1}\right]}\right),  \quad k \in \{0,1\}
\end{eqnarray}
$\beta = 1/T$ is the inverse temperature of the system, $E^0_{\left[\sigma_{i:i+1}\right]} \left( E^1_{\left[\sigma_{i:i+1}\right]}\right)$ is the lattice energy of the spins between columns $i$ and $i+1$ when the diagonal bonds connect the top (bottom) left and bottom (top) right spin-lattice sites and $T^k_{\left[\sigma_{i}\right], \left[\sigma_{i+1}\right]}$  is the transfer matrix whose rows and columns represent all the possible combinations of spins at column $i$ $\left(\left[\sigma_{i}\right] \right)$ and column $i + 1$ $\left(\left[\sigma_{i+1}\right]\right)$ respectively.  
We redefine the total partition function ($Z_\beta$) of the spin system as:  
\begin{eqnarray}
    Z_\beta &=&\sum_{\left[\sigma_{0:L-1}\right]}\exp \left(-\beta \mathcal{H}\left(\sigma\right) \right) = \text{Tr}\left[T^{\lfloor L/2\rfloor}\right] 
\end{eqnarray}
With the partition function $\left(Z_\beta \right)$ of the lattice, we then calculate the expectation of one and two spin operators:
\begin{eqnarray}
    \langle \sigma_{2i} \sigma_{2i + 2j} \rangle &=& \lim_{\beta \rightarrow \infty} \lim_{L \rightarrow \infty} \frac{1}{Z_\beta} \text{Tr}\biggl[T^i\tau^z T^j\nonumber\\
    & & \tau^z T^{\lfloor L/2\rfloor - i - j}\biggr] \\
    \langle \sigma_{2i} \rangle &=&   \lim_{\beta \rightarrow \infty} \lim_{L \rightarrow \infty} \frac{1}{Z_\beta} \text{Tr}\biggl[T^i\tau^z T^{\lfloor L/2\rfloor - i} \biggr] 
\end{eqnarray}
 $\tau^z$ is the Pauli-Z matrix and $T = T^0T^1$. These spin expectation values can then be used to calculate the ground state energy and  the spin correlation. \textcolor{red}{A more detailed description is provided in the Supplementary Information}

\begin{figure}
\includegraphics[width=0.4\textwidth]{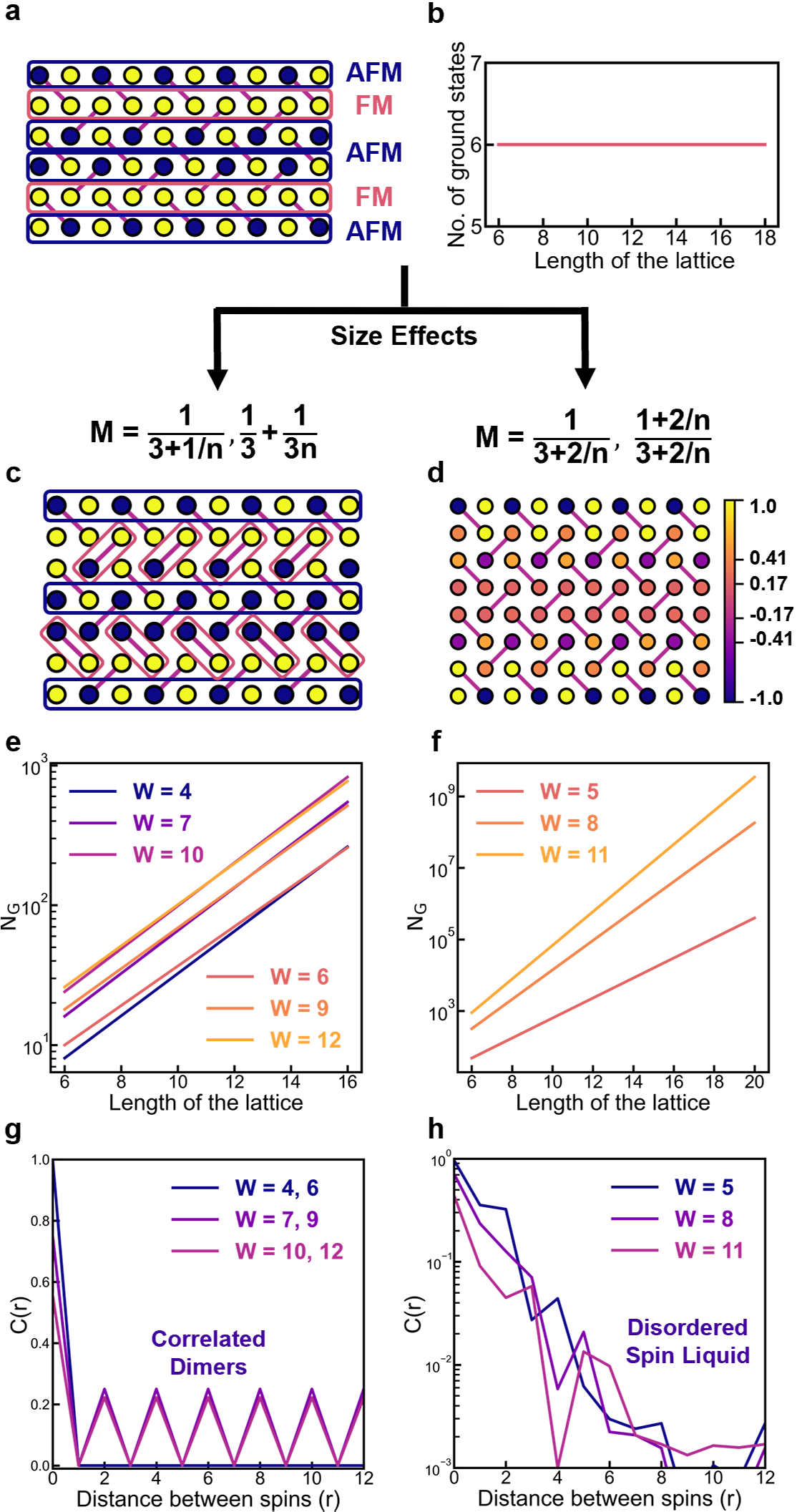}
\caption{\textbf{Characterisation of classical spin liquid phase:} (\textbf{a}) The ordered UUD (1/3) magnetization spin state in the thermodynamic limit. (\textbf{b}) The number of degenerate ground states for UUD is independent of the size of the system. Size effect introduces two types of spin liquid depending on the width of the lattice: (\textbf{c}) classical spin liquid for widths $3n$ and $3n + 1$ where the correlated dimers in pink boxes are sandwiched between antiferromagnetic rows and (\textbf{d}) disordered classical spin liquid for width $3n + 2$ where each spin is fractionally spin-spin correlated to all its neighboring spins. The color bar represents the average magnetization for each spin site. (\textbf{e}), (\textbf{f}) Both spin states are identified as classical spin liquids via the exponential increase of ground states with the length of the system. (\textbf{g}) Long-range spin-spin correlation is present along a row of the lattice in the correlated dimer spin liquid state. (\textbf{h}) Exponential spin-spin correlation decay in the disordered spin liquid state.}
\label{Fig3}
\end{figure}

As the transfer matrix for our studied lattice is non-diagonalizable, we resort to the numerical estimation of the partition function and the required observables.  
We choose a large enough length $(L \geq 20$) and low enough temperature $(\beta \geq 15$) so that the calculated observables converge to their thermodynamic limits (\textcolor{red}{Fig. S5}). 
To study size effects, we introduce periodic boundary conditions along the transverse direction (length) and open boundary conditions along the lateral direction (width) of the lattice.
Subsequently, we use the numerical estimation of the partition function to obtain phase diagrams by varying the interaction bond ratio, $J_2/J_1$ from 0 to 3, and the applied magnetic field, $h_z$ from -8 to 8 for widths $W$ varying from 4 to 12.
The above parameter ranges are chosen to ensure the presence of $1/3$ magnetization plateau, according to studies on the SS lattice \cite{Kairys2020, Dublenych2012}.
The ground state spin configuration for every magnetization phase is deduced from the spin-spin correlation to all its nearest neighbors and the average magnetization spin site calculations. 
These ground states are also verified by comparing their energies to the ground state energies obtained by Monte-Carlo sampling the Boltzmann distribution of the Shastry-Sutherland Ising lattice \cite{Kotze2008, Huang2012}.
To further characterize the ground states of the laterally confined SS Ising lattice, we calculate the spatial variation of spin-spin correlation along the length of the lattice and the number of degenerate ground states (as shown in Fig.\ref{Fig3}e and Fig.\ref{Fig3}f) using the following equations: 

\begin{eqnarray}
    F &=& -k_bT \ln\left( Z_\beta\right)  \\
    S &=& \frac{E_{gs} - F}{T} = k_b \ln (N_G) \label{S_eqn}
\end{eqnarray}

$k_b$ is the Boltzmann constant, $F$ is the free energy, $S$ is the entropy, $E_{gs}$ is the ground state energy and $N_G$ is the number of degenerate ground states.
For calculating $N_G$, we choose low enough temperatures such that the average energy of the system converges to the ground state energy ($\langle E \rangle \approx E_{gs}$) and every spin configuration other than the ground state has negligible Boltzmann probabilities.
Thus, the entropy $S$ takes the form of eq.(\ref{S_eqn}).
It should be noted that macroscopically degenerate systems (the number of degenerate ground states is a macroscopic property) have non-trivial entropy even in the zero temperature limit \cite{Balents2010}.

Our main results predict the emergence of a classical spin liquid phase from an ordered magnetic state when the Shastry-Sutherland (SS) Ising lattice is confined along the lateral direction. 
Fig.\ref{Fig2}a shows the magnetization phase diagram of the SS lattice in the thermodynamic limit. 
The ferromagnetic (FM), antiferromagnetic (AFM), and 1/3 fractional magnetic phases have a finite number of degenerate ground states while the dimer magnetic phase is macroscopically degenerate at the zero temperature limit, due to the uncorrelated antiferromagnetic dimers present along the diagonal bonds of the SS lattice shown in Fig.\ref{Fig2}a.
But our study focuses on the $1/3$ fractional phase containing 6 degenerate ground states where each triangle in the lattice contains two up spins and one down spin (UUD  spin configuration) \cite{Dublenych2012}.
With size effects introduced by confining the lattice along the lateral direction, the ordered  1/3 fractional magnetization phase gets fragmented into slightly different valued magnetic phases depending upon the width of the lattice. In contrast, the other magnetization phases (FM, AFM, and Dimer) retain their shapes and magnetization values as shown in Fig.\ref{Fig2}. 
We categorize the phase diagrams of the laterally confined lattice according to their widths: $W = 3n$, $W = 3n + 1$, and $W = 3n + 2$, which also determines the region of the 1/3 fractional phase that transforms into the classical spin liquid phase, identified by its finite temperature-entropy product in the zero temperature limit. 
Below a critical width, in (i) $W = 3n$, the region of the 1/3 fractional phase along the borders of the FM and within a longitudinal field ($h_z$) range transforms into the $1/3 + 1/3n$ classical spin liquid phase, in (ii) $W = 3n + 1$, the region of the 1/3 fractional phase bordering the dimer phase where $J_2 \geq 2 J_1$ transforms into the $1/(3 + 1/n)$ classical spin liquid phase and finally, in (iii) $W = 3n + 2$, the thin region of the 1/3 fractional phase along the FM and the dimer phase transforms into the $(1 + 2/n)/(3 + 2/n)$ and the $1/(3 + 2/n)$ classical spin liquid phase respectively. Fig. \ref{Fig2}b shows all the above emergent classical spin liquid phases as the black hashed region.

The ground state spin configurations of the aforementioned classical spin liquid phases can be grouped into two categories, as shown in Fig.\ref{Fig3}. 
Both $W=3n$ and $W=3n + 1$ have the same ground state spin liquid phase described by rows of correlated dimers sandwiched between antiferromagnetic rows (Fig.\ref{Fig3}(c)). 
Due to the exponential combinations possible among the correlated dimers, the number of degenerate ground states is a macroscopic property similar to the dimer phase (Fig.\ref{Fig3}(e)) \cite{Dublenych2012}.
Lattices with small widths ($W=4,6$) contain uncorrelated dimers whereas for larger widths, there exist persistent long-range correlated dimers (Fig.\ref{Fig3}(g)). 
For $W=3n+2$, the classical spin liquid phase is entirely disordered, characterized by the average fractional magnetization per site and the fractional spin-spin correlation among all its nearest and second-nearest neighbors (Fig.\ref{Fig3}(d)). 
Similar to the previous spin liquid phase, the number of degenerate ground states is a macroscopic property (Fig.\ref{Fig3}(f)), but in contrast to the long-range spin-spin correlation present in the previous case,  the spin-spin correlation decays exponentially along the length of the lattice for this particular spin liquid state (Fig.\ref{Fig3}(h)).
The detailed magnetization phase diagrams and the ground states of all the magnetization plateaus for widths varying from $4-12$ are given in the supplemental \textcolor{red}{Fig.S1 - S4}.
The emergence of these exotic phases is explained by the lateral confinement or size effects introducing additional frustration in the bulk of the lattice where the spin arrangement and interaction energy at the edges destroy the long-range spin order in the bulk and reduces the total energy of the system \cite{Maric2020}. 
Thus, these results demonstrate the key role of size effects in stabilizing spin liquids.  

The dimers in the spin liquid states of lattice widths $3n$ and $3n + 1$ are always sandwiched between antiferromagnetic rows of opposite spin arrangements.
Thus, a lattice structure must have width $3d + 1$ to contain d rows of dimers interleaved with d + 1 antiferromagnetic rows.
This spin arrangement also leads to AFM spin configuration at the edges, which is favorable for low hz values, stabilizing the dimer spin liquid for $W = 3n + 1$. 
In contrast, for the higher hz values, the spins along both edges favor the FM configuration to lower the ground state energy compared to the FM and AFM spin edge configuration in the bulk UUD configuration.
Adding the two FM spin rows at the edges to the previous $d$ rowed dimer structure gets us a total of $3d + 3 = 3n$, thus leading to the spin liquid stabilization for $W = 3n$ at higher hz values.
But for lattices with width, $3n + 2$ can only support the $d$-dimer structure with an FM and an AFM spin configuration at the edges, leading to an increase in the energy for lower and higher hz values. 
This leads to the formation of a new spin liquid with no dimers and exponential decaying spin correlation, with both the edges having FM spin configurations for higher hz values and AFM spin configurations for the lower hz values (\textcolor{red}{Fig. S4}).

In summary, a classical spin liquid state emerges from the ordered UUD magnetization state in a laterally confined classical SS Ising lattice. In particular, two classical spin liquid states are observed depending on the width of the lattice: (i) correlated dimers sandwiched between antiferromagnetic rows (widths $3n$ and $3n + 1$), and (ii) spin liquid state with fractional correlation to all its neighboring interacting spins (width $3n + 2$). 
This novel physical phenomenon originates from spin arrangement at the edges which destroy the long-range bulk order to reduce the total energy of the lattice; therefore, size effects and boundary conditions underlie spin liquid stabilization in laterally confined frustrated 2D lattices. 
%
%
This approach can pave the way for stabilizing new spin liquid systems through lateral confinement of quasi-two-dimensional magnetic materials, e.g. rare-earth tetraborides in which the ionic arrangement is topologically equivalent to the SS lattice \cite{Yoshii2008, Kim2009, Trinh2018, Siemensmeyer2008}.


This work was supported by ASCENT, one of six centers in JUMP, a Semiconductor Research Corporation (SRC) program sponsored by DARPA.

\newpage
\bibliography{main}

%
%
%
%
%
%
%





\onecolumngrid

\pagebreak
\begin{center}
\textbf{\large Supplemental Materials: Emergent Classical Spin Liquid Phases in an Ising Lattice via Size Effects}
\end{center}

\setcounter{equation}{0}
\setcounter{figure}{0}
\setcounter{table}{0}
\setcounter{page}{1}
\makeatletter
\renewcommand{\theequation}{S\arabic{equation}}
\renewcommand{\thefigure}{S\arabic{figure}}
\renewcommand{\bibnumfmt}[1]{[S#1]}
\renewcommand{\citenumfont}[1]{S#1}

\section{Calculation of Partition Function}

Using the total partition function, we calculate the ground state energy, ground state spin configuration, spin-spin correlation, and magnetization phase diagram of the SS Ising lattice.
We assume the size of the lattice to be $W \times L$ and define variables (i) $\sigma_i^j$, the spin site at column $i$ and row $j$, (ii) $\left[\sigma_{a:b}\right]$, the set of all spin sites between column $a$ and column $b$.
Using the above definitions, we introduce the following variables:
\begin{eqnarray}
E^H_{\left[\sigma_{i:i+1}\right]} &=& J_1 \sum_{j = 0}^{W-1} \sigma_i^j \sigma_{i+1}^j\\
E^V_{\left[\sigma_{i:i+1}\right]} &=& J_1 \sum_{j = 0}^{W-1} \sigma_i^j \sigma_i^{j + 1} + \sigma_{i+1}^j \sigma_{i+1}^{j+1}\\
E^k_{\left[\sigma_{i:i+1}\right]} &=& E^H_{\left[\sigma_{i:i+1}\right]} + \frac{1}{2}E^V_{\left[\sigma_{i:i+1}\right]} + 
 J_2\sum_{j=0}^{\lfloor W/2 \rfloor} \sigma_{i}^{2j}\sigma_{i + 1}^{2j + (-1)^k} \quad ,k \in \{0,1\}\\
 T^k_{\left[\sigma_{i}\right], \left[\sigma_{i+1}\right]} &=& \exp \left(- \beta E^k_{\left[\sigma_{i:i+1}\right]}\right)
\end{eqnarray}
$\beta = 1/T$ is the inverse temperature of the system, $E^0_{\left[\sigma_{i:i+1}\right]} \left( E^1_{\left[\sigma_{i:i+1}\right]}\right)$ is the lattice energy of the spins between columns $i$ and $i+1$ when the diagonal bonds connect the top (bottom) left and bottom (top) right spin-lattice sites and $T^k_{\left[\sigma_{i}\right], \left[\sigma_{i+1}\right]}$  is the transfer matrix whose rows and columns represent all the possible combinations of spins at column $i$ $\left(\left[\sigma_{i}\right] \right)$ and column $i + 1$ $\left(\left[\sigma_{i+1}\right]\right)$ respectively.  
We redefine the total partition function ($Z_\beta$) of the spin system as:  
\begin{eqnarray}
    Z_\beta &=&\sum_{\left[\sigma_{0:L-1}\right]}\exp \left(-\beta \mathcal{H}\left(\sigma\right) \right) \nonumber\\ 
    &=& \sum_{\left[\sigma_{0:L-1}\right] } \exp \biggl(-\beta \sum_{i=0}^{\lfloor L/2 \rfloor} E^0_{\left[ \sigma_{2i:2i+1}\right]} 
    + E^1_{\left[ \sigma_{2i+1:2i+2}\right]} \biggr)\\
    Z_\beta &=& \sum_{\left[\sigma_{0:L-1}\right]} \prod_{i=0}^{\lfloor L/2\rfloor} T^0_{\left[\sigma_{2i}\right], \left[\sigma_{2i+1}\right]}  T^1_{\left[\sigma_{2i+1}\right], \left[\sigma_{2i+2}\right]}\\
    Z_\beta &=& \text{Tr}\left[\left(T^0 T^1 \right)^{\lfloor L/2\rfloor}\right] = \text{Tr}\left[T^{\lfloor L/2\rfloor}\right]
\end{eqnarray}
With the partition function $\left(Z_\beta \right)$ of the lattice, we then calculate the expectation of one and two spin operators:
\begin{eqnarray}
    \langle \sigma_{2i} \sigma_{2i + 2j} \rangle &=& \lim_{\beta \rightarrow \infty} \lim_{L \rightarrow \infty} \frac{1}{Z_\beta} \text{Tr}\biggl[T^i\tau^z T^j  \tau^z T^{\lfloor L/2\rfloor - i - j}\biggr] \\
    \langle \sigma_{2i} \rangle &=&   \lim_{\beta \rightarrow \infty} \lim_{L \rightarrow \infty} \frac{1}{Z_\beta} \text{Tr}\biggl[T^i\tau^z T^{\lfloor L/2\rfloor - i} \biggr] 
\end{eqnarray}
 $\tau^z$ is the Pauli-Z matrix. These spin expectation values can then be used to calculate the ground state energy and  the spin correlation:
 \begin{eqnarray}
    C(2j) &=& \langle \sigma_{2i} \sigma_{2i + 2j} \rangle - \langle \sigma_{2i} \rangle \langle \sigma_{2i + 2j}  \rangle \\
    E_{gs} &=& J_1\sum_{\langle i,j \rangle} \langle \sigma_i\sigma_j \rangle + J_2\sum_{\langle \langle i,j \rangle \rangle} \langle \sigma_i\sigma_j \rangle   + h\sum_i \langle\sigma_i\rangle
\end{eqnarray}
$E_{gs}$ is the ground state energy and $C(2j)$ is the spin-spin correlation between two spins at lattice sites $2j$ distance apart.

\newpage 
\section{Phase Diagrams and Ground State Lattices}

\begin{figure*}[h]
\includegraphics[width=0.95\columnwidth]{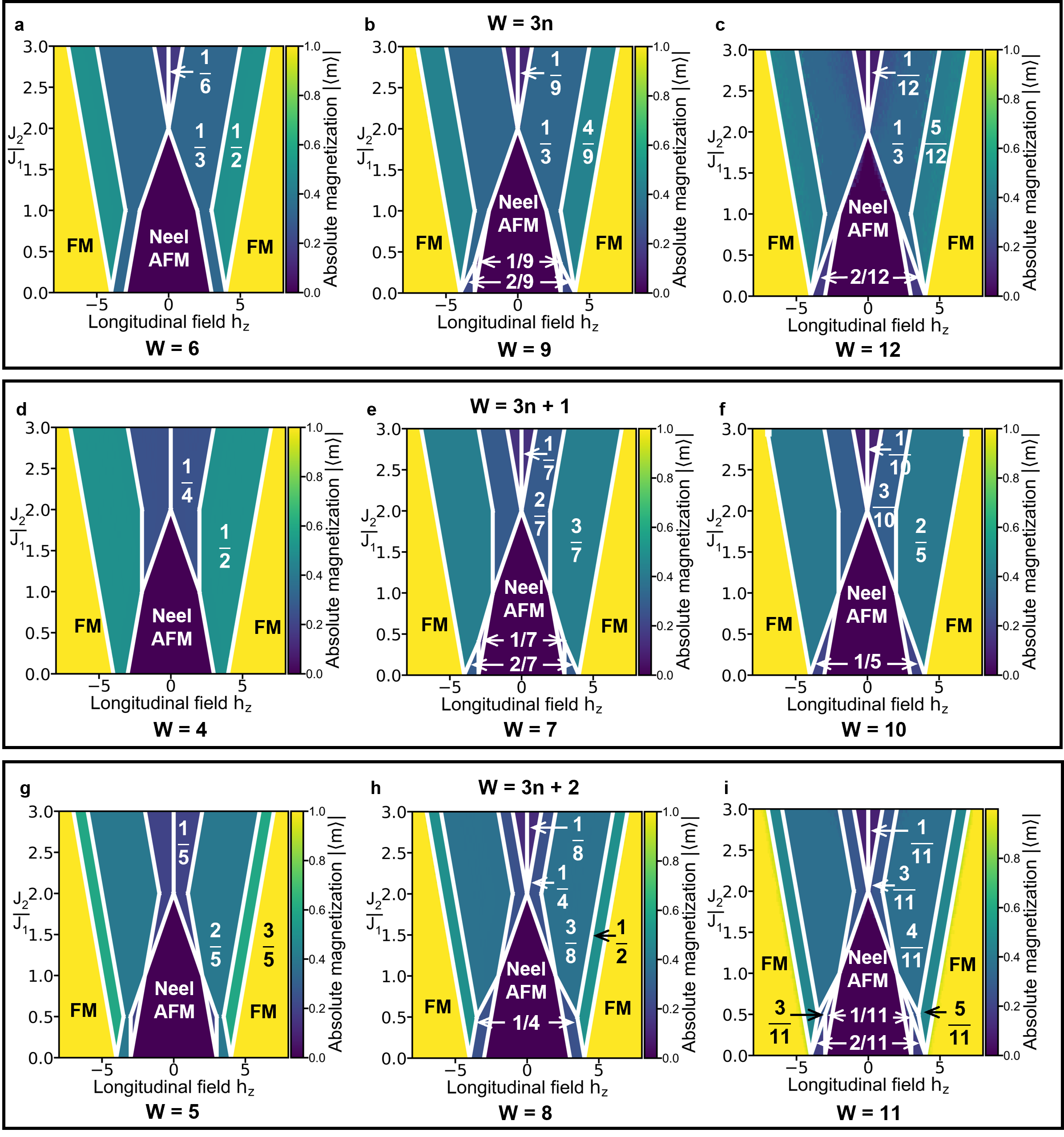}
\caption{\textbf{Magnetization phase diagrams for widths 4 to 12:} The phase diagrams have been grouped according to modulus 3 of the width of the lattice. The following phases were not shown in the main Fig.\ref{Fig2} due to brevity : (i) M = 1/9 for W = 9, (ii) M = 1/7 for W = 7, (iii) M = 1/11, M = 3/11 for W = 11, (iv) M = 1/5 for W = 5. 
}
\label{SuppFig1}
\end{figure*}
\vfill

\vfill
\begin{figure*}
\includegraphics[width=0.95\columnwidth]{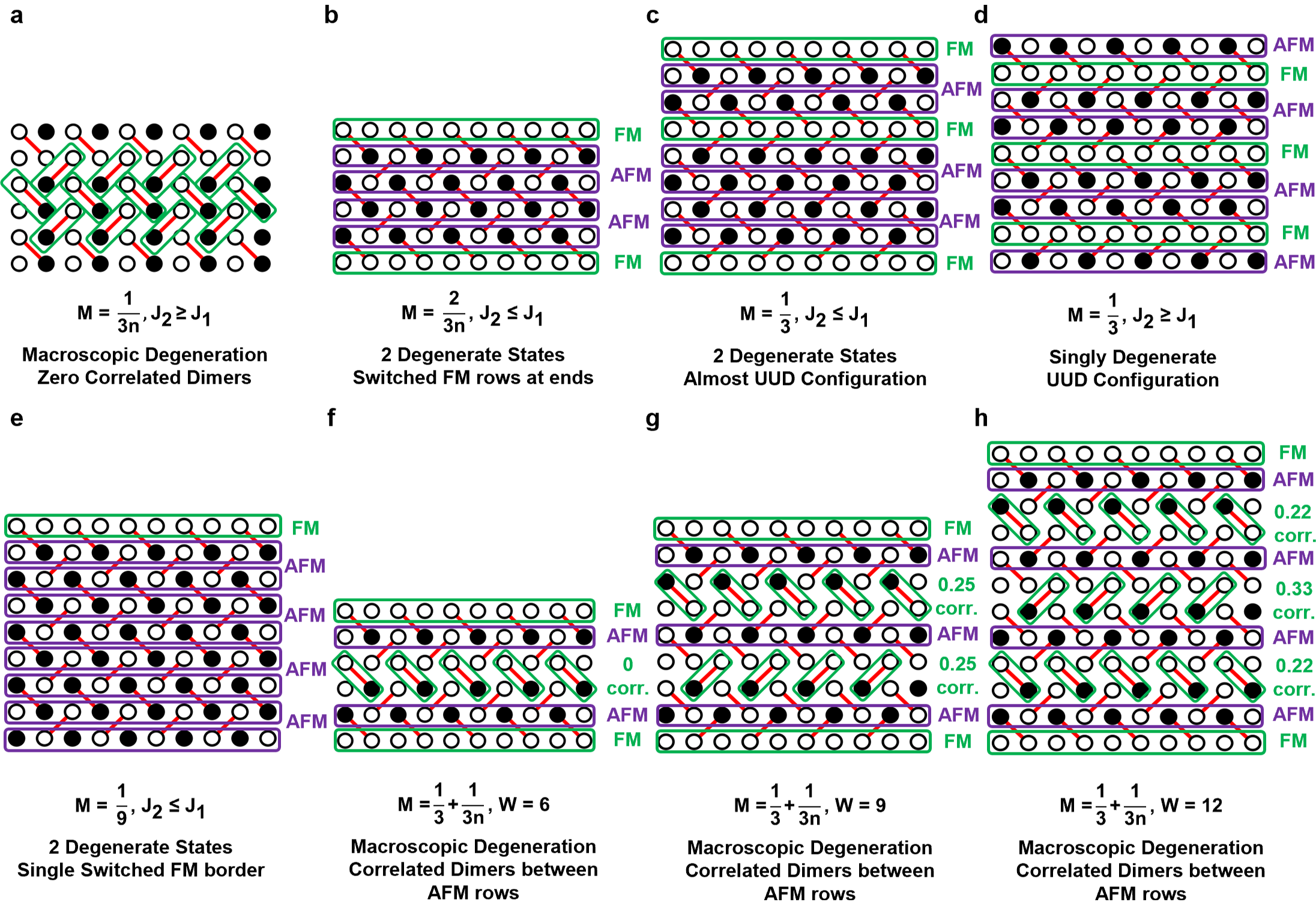}
\caption{\textbf{Ground state solutions of all magnetization phases for width $3n$: } The description of the spin configuration and the number of degenerate ground states are given below each lattice diagram. The green boxes represent the antiferromagnetic dimers, the white circle represents +1 and the black circle represents -1. (\textbf{a}) is the original dimer phase in the thermodynamic limit with antiferromagnetic edges, (\textbf{b})-(\textbf{e}) are the ordered spin lattice states and (\textbf{f})-(\textbf{h}) are the classical spin liquids. The correlation among the dimers is written beside each dimer row.}
\label{SuppFig2}
\end{figure*}
\vfill

\vfill
\begin{figure*}
\includegraphics[width=0.95\columnwidth]{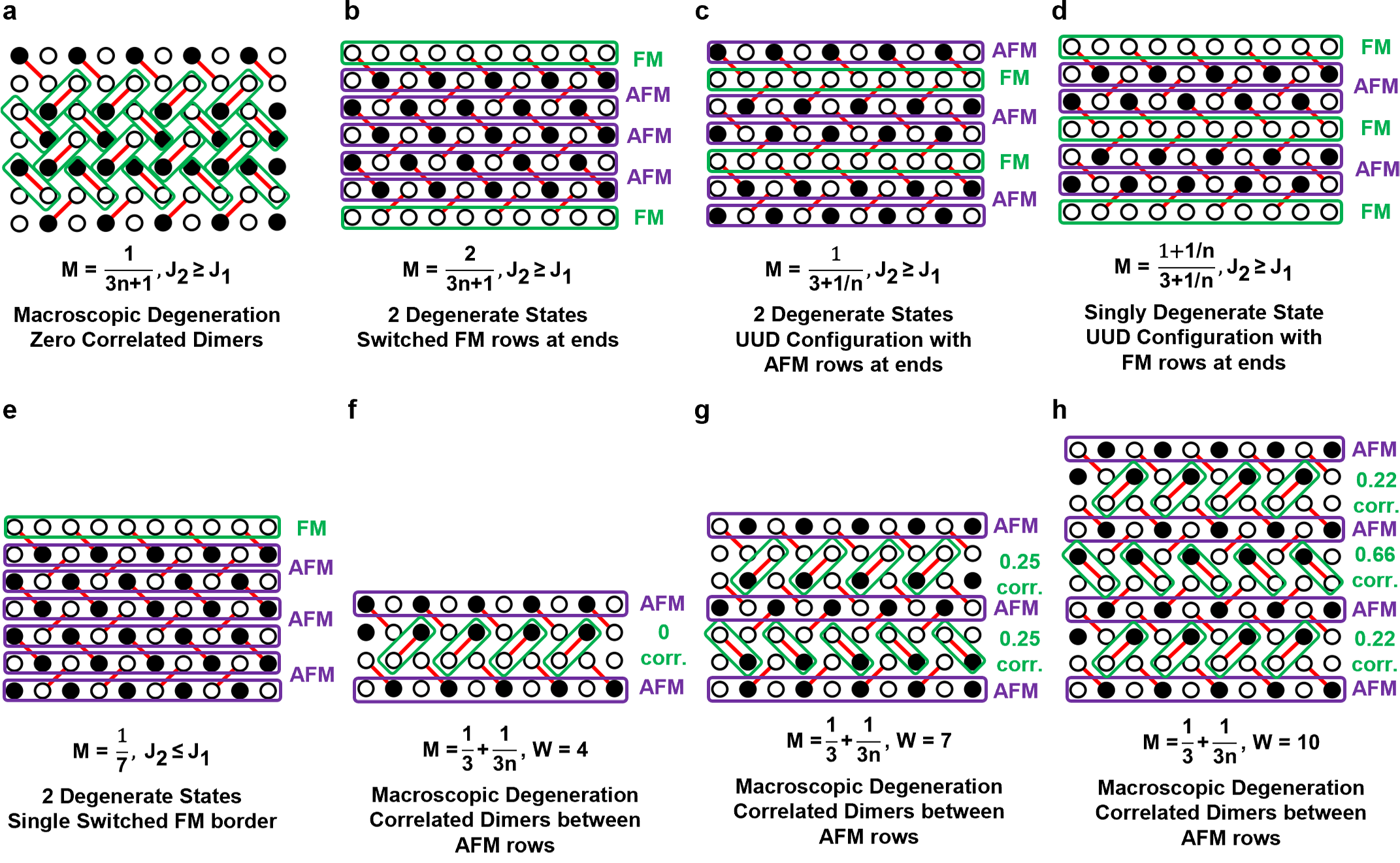}
\caption{\textbf{Ground state solutions of all magnetization phases for width $3n + 1$:}  The description of the spin configuration and the number of degenerate ground states are given below each lattice diagram. The green boxes represent the antiferromagnetic dimers, the white circle represents +1 and the black circle represents -1. (\textbf{a}) is the original dimer phase in the thermodynamic limit with antiferromagnetic edges, (\textbf{b})-(\textbf{e}) are the ordered spin lattice states and (\textbf{f})-(\textbf{h}) are the classical spin liquids. The correlation among the dimers is written beside each dimer row.}
\label{SuppFig3}
\end{figure*}
\vfill

\vfill
\begin{figure*}
\includegraphics[width=0.95\columnwidth]{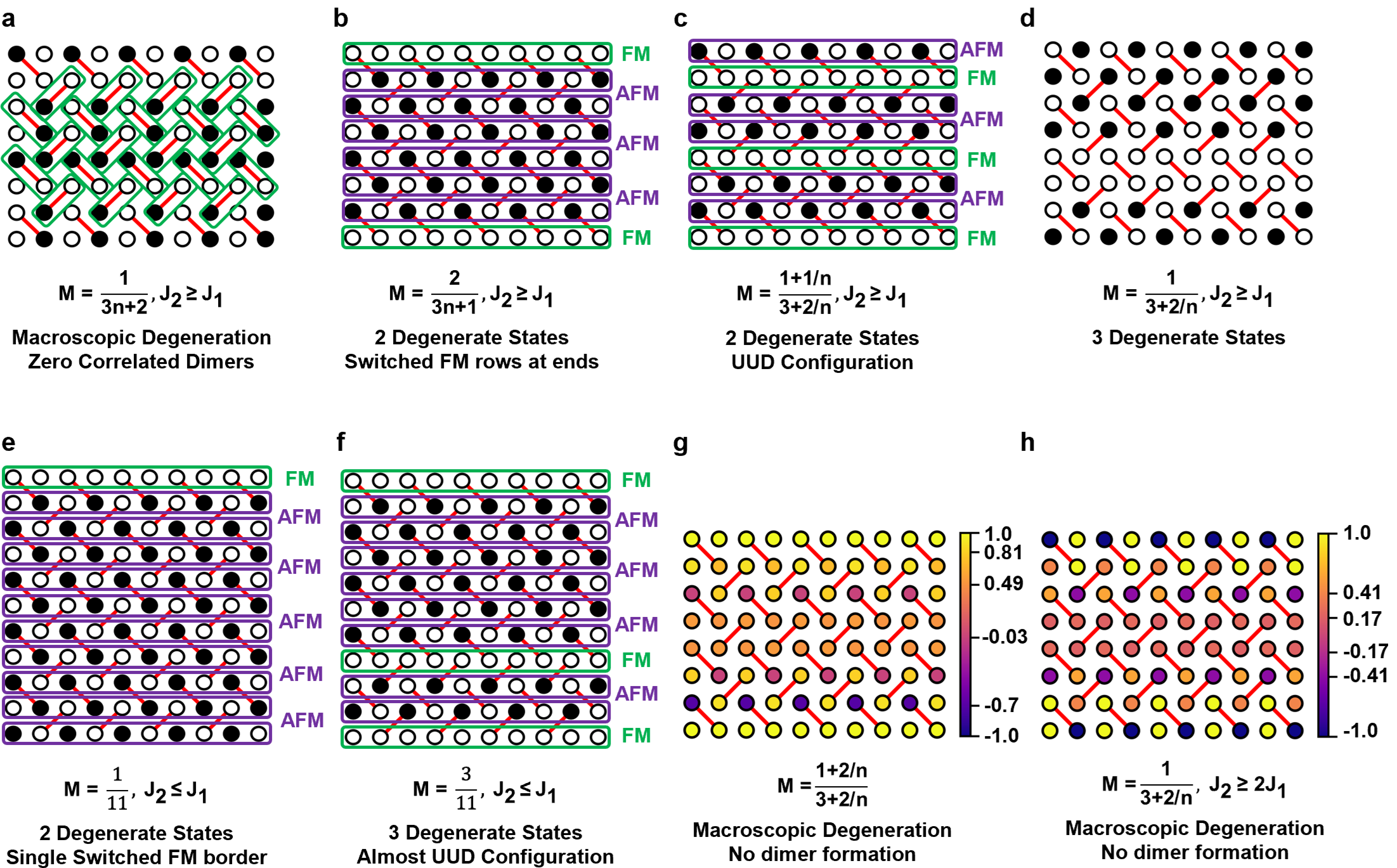}
\caption{\textbf{Ground state solutions of all magnetization phases for width $3n + 2$:} The description of the spin configuration and the number of degenerate ground states are given below each lattice diagram. The green boxes represent the antiferromagnetic dimers, the white circle represents +1 and the black circle represents -1. (\textbf{a}) is the original dimer phase in the thermodynamic limit with antiferromagnetic edges, (\textbf{b})-(\textbf{f}) are the ordered spin lattice states and (\textbf{g})-(\textbf{h}) are the disordered classical spin liquids due to the fractional spin-spin correlations among all the interacting spins. The color bar represents the average magnetization for each spin site.}
\label{SuppFig4}
\end{figure*}
\vfill

\vfill
\begin{figure}
\includegraphics[width=0.5\columnwidth]{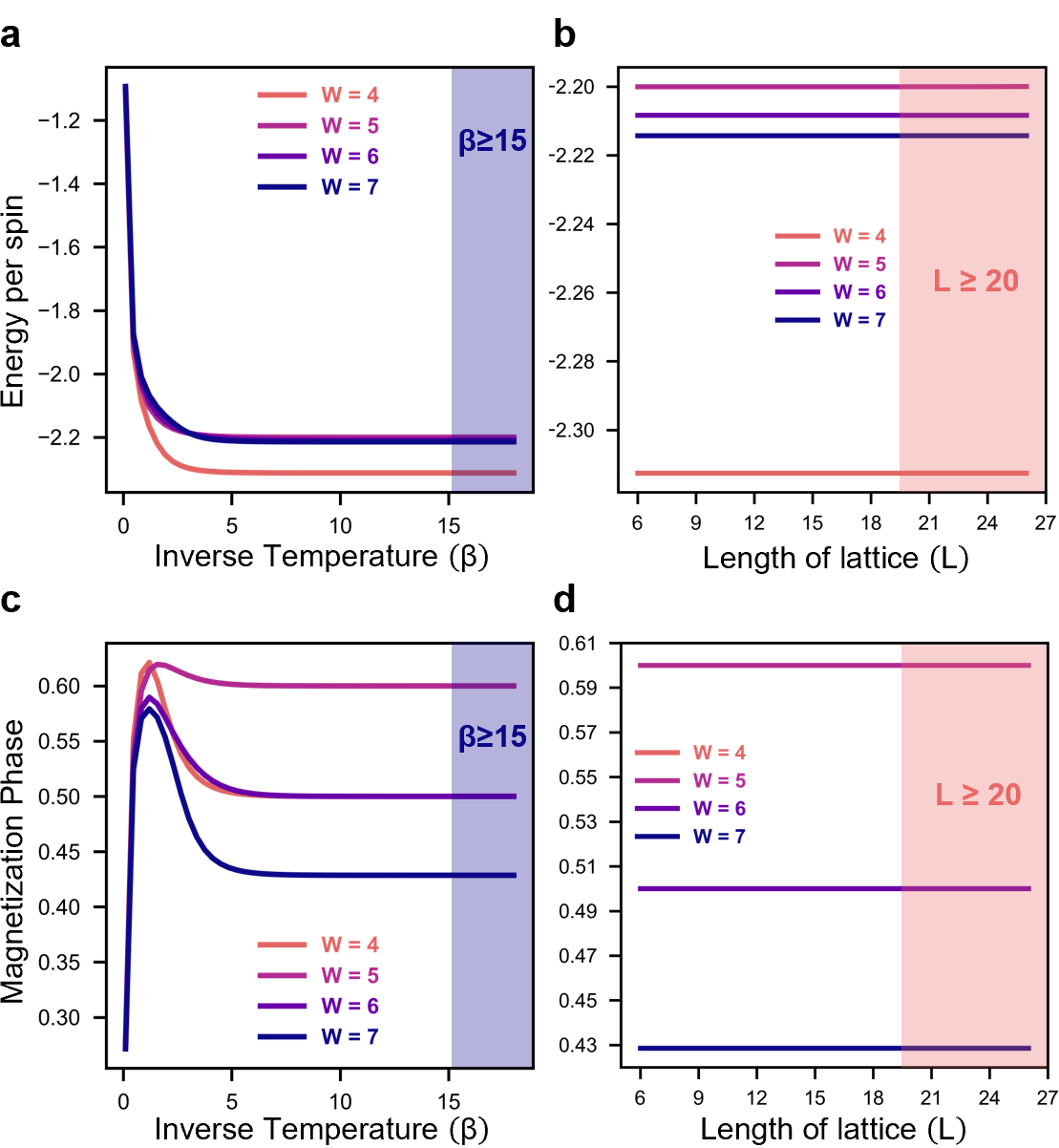}
\caption{\textbf{Convergence to the thermodynamic value}: (\textbf{a}) Energy per spin converges to the thermodynamic value with an increase in inverse temperature ($\beta$). Therefore, we choose $\beta >= 15$ for all computations. (\textbf{b}) Energy per spin remains almost constant with an increase in the length of the lattice. We choose $L >= 20$ to ensure thermodynamic convergence. (\textbf{c}) The magnetization phase converges to the thermodynamic limit with an increase in inverse temperature ($\beta$). Therefore, we choose $\beta >= 15$ for all computations. (\textbf{d}) The magnetization phase remains almost constant with an increase in the length of the lattice. We choose $L >= 20$ to ensure thermodynamic convergence. }
\label{SuppFig5}
\end{figure}
\vfill

\end{document}